\documentclass[10pt, conference, twocolumn, letterpaper]{IEEEtran}
\IEEEoverridecommandlockouts
\usepackage{cite}
\usepackage{amsmath,amssymb,amsfonts}
\usepackage{algorithmic}
\usepackage{graphicx}
\usepackage{textcomp}
\usepackage{xcolor}
\usepackage{url}
\def\BibTeX{{\rm B\kern-.05em{\sc i\kern-.025em b}\kern-.08em
    T\kern-.1667em\lower.7ex\hbox{E}\kern-.125emX}}
\begin{document}

\title{Accelerate Intermittent Deep Inference
}

\author{\IEEEauthorblockN{Ziliang Zhang}
University of California, Riverside \\ ziliang.zhang@email.ucr.edu \\
}

\maketitle
\textbf{Emerging research in edge devices and micro-controller units (MCU) enables on-device computation of Deep Learning Training and Inferencing\footnote{In this paper we will use Deep Learning Inferencing and Deep Inference/Deep Inferencing interchangeably} tasks. More recently, contemporary trends focus on making the Deep Neural Net (DNN) Models runnable on battery-less intermittent devices. One of the approaches is to shrink the DNN models by enabling weight sharing, pruning, and conducted Neural Architecture Search (NAS) with optimized search space to target specific edge devices \cite{Cai2019OnceFA} \cite{Lin2020MCUNetTD} \cite{Lin2021MCUNetV2MP} \cite{Lin2022OnDeviceTU}. Another approach analyzes the intermittent execution and designs the corresponding system by performing NAS that is aware of intermittent execution cycles and resource constraints \cite{iNAS} \cite{HW-NAS} \cite{iLearn}.}

\textbf{However, the optimized NAS was only considering consecutive execution with no power loss, and intermittent execution designs only focused on balancing data reuse and costs related to intermittent inference and often with low accuracy. We proposed Accelerated Intermittent Deep Inference to harness the power of optimized inferencing DNN models specifically targeting SRAM under 256KB and make it schedulable and runnable within intermittent power. Our main contribution is: (1) Schedule tasks performed by on-device inferencing into intermittent execution cycles and optimize for latency; (2) Develop a system that can satisfy the end-to-end latency while achieving a much higher accuracy compared to baseline \cite{iNAS} \cite{HW-NAS}}

\begin{IEEEkeywords}
Emebedded System, Neural Network, Intermittent Device
\end{IEEEkeywords}

\section{Introduction}
Contemporary research in edge devices and micro-controller units (MCUs) is enabling on-device computation of Deep Learning training and inference tasks. Specifically, recent trends have focused on making Deep Neural Net (DNN) models run on battery-less intermittent devices.

To achieve this, researchers have proposed two approaches. The first involves shrinking DNN models by enabling weight sharing, pruning, and conducting Neural Architecture Search (NAS) with optimized search space to target specific edge devices. The second approach involves analyzing intermittent execution and designing the corresponding system by performing NAS that is aware of intermittent execution cycles and resource constraints.

Energy harvesting has become a popular solution for edge devices, as it uses ambient energy instead of batteries, making it more sustainable and cost-effective. However, devices that rely on energy harvesting often suffer from power failures and have to execute intermittently. Communicating with remote servers requires significantly more energy than local computation or sensing, which has led to the development of on-device intelligence and the execution of deep neural network (DNN) inference on intermittent systems. Neural architecture search (NAS) techniques have been developed to automatically find highly accurate neural networks that can efficiently execute on deployed systems. With the increasing demand for deployment on battery-less edge devices, intermittent-aware neural architecture search is becoming crucial.

DNN inference under intermittent power requires accumulative execution across power cycles, as ambient power is typically unstable and too weak for continuous execution. Even highly power-efficient hardware may take multiple power cycles to complete a full inference. Therefore, intermittent inference execution is necessary on lightweight energy-harvesting devices, and in most cases, it cannot be avoided by using power-efficient hardware. Intermittent execution behavior is significantly different from continuous execution behavior. Inference must execute safely within an energy budget in each power cycle, and intermittent inference approaches ensure that progress preservation during inference does not impede performance and guarantee correct progress recovery to resume inference upon power resumption.

Neural architecture search is a design-time process that explores the neural architecture space to find DNNs with high accuracy for a particular application/dataset. Recent research has proposed hardware-aware NAS approaches, which co-explore both the neural architecture space and execution design space, to find DNNs with high accuracy and associated execution designs that do not violate a target inference latency requirement. However, existing NAS approaches assume the deployed system executes under continuous power, which is not suitable for intermittently-powered inference systems.

This work is motivated by the unsuitability of existing NAS approaches for intermittently-powered inference systems. NAS primarily seeks to maximize data reuse without being aware of intermittent execution behavior, which may result in an inefficient and unsafe neural network and associated execution design when deployed on an intermittent inference system. Intermittent-aware neural architecture search should find the right balance between data reuse and the costs related to progress preservation and recovery, while ensuring the power-cycle energy budget is not exceeded.

However, previous optimized NAS solutions were only considering consecutive execution with no power loss, and intermittent execution designs only focused on balancing data reuse and costs related to intermittent inference, often with low accuracy. Thus, the authors propose a new solution to harness the power of optimized inferencing DNN models specifically targeting SRAM under 256KB and making it schedulable and runnable within intermittent power.

The design challenge would involve scheduling tasks performed by on-device inference into intermittent execution cycles and optimizing for latency, while also achieving higher accuracy compared to the baseline. This would require developing a system that can satisfy the end-to-end latency requirements while maintaining a high level of accuracy.

\section{Background}
\subsection{Intermittent DNN}

Deep neural networks (DNNs) are a class of artificial neural networks that have multiple layers of computation between input and output. These layers are arranged sequentially and are composed of different types of layers such as convolutional, pooling, and fully connected layers. In a DNN, the output of one layer is fed as input to the subsequent layer, and this process continues until the output is produced. Each layer's architecture directly affects the accuracy and inference latency of the DNN model.

\begin{figure}[htbp]
  \centering
  \includegraphics[width=0.95\linewidth]{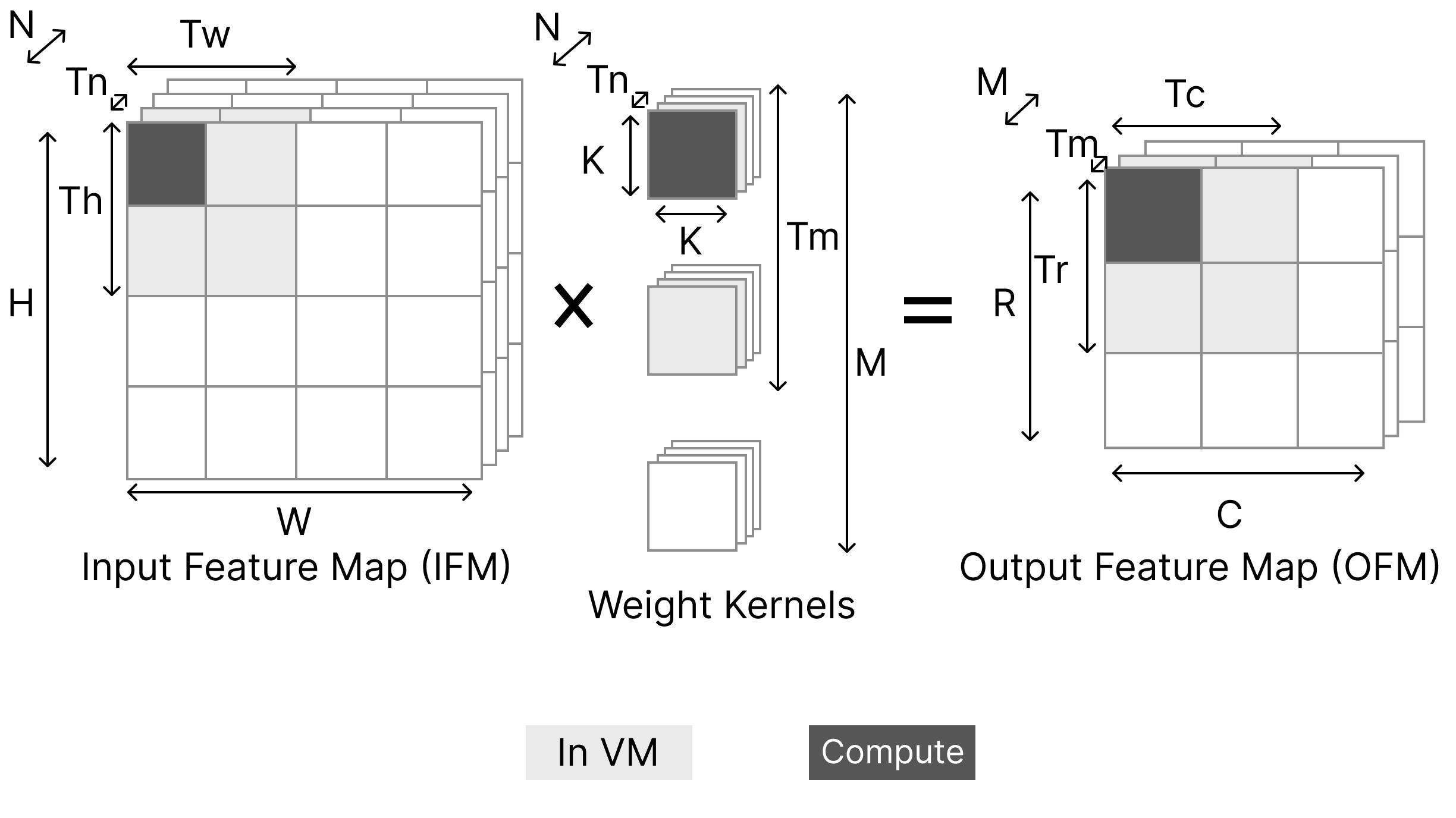}
  \caption{Intermittent DNN Execution Pattern}
  \label{dnn}
\end{figure}

\begin{figure}[htbp]
  \centering
  \includegraphics[width=0.95\linewidth]{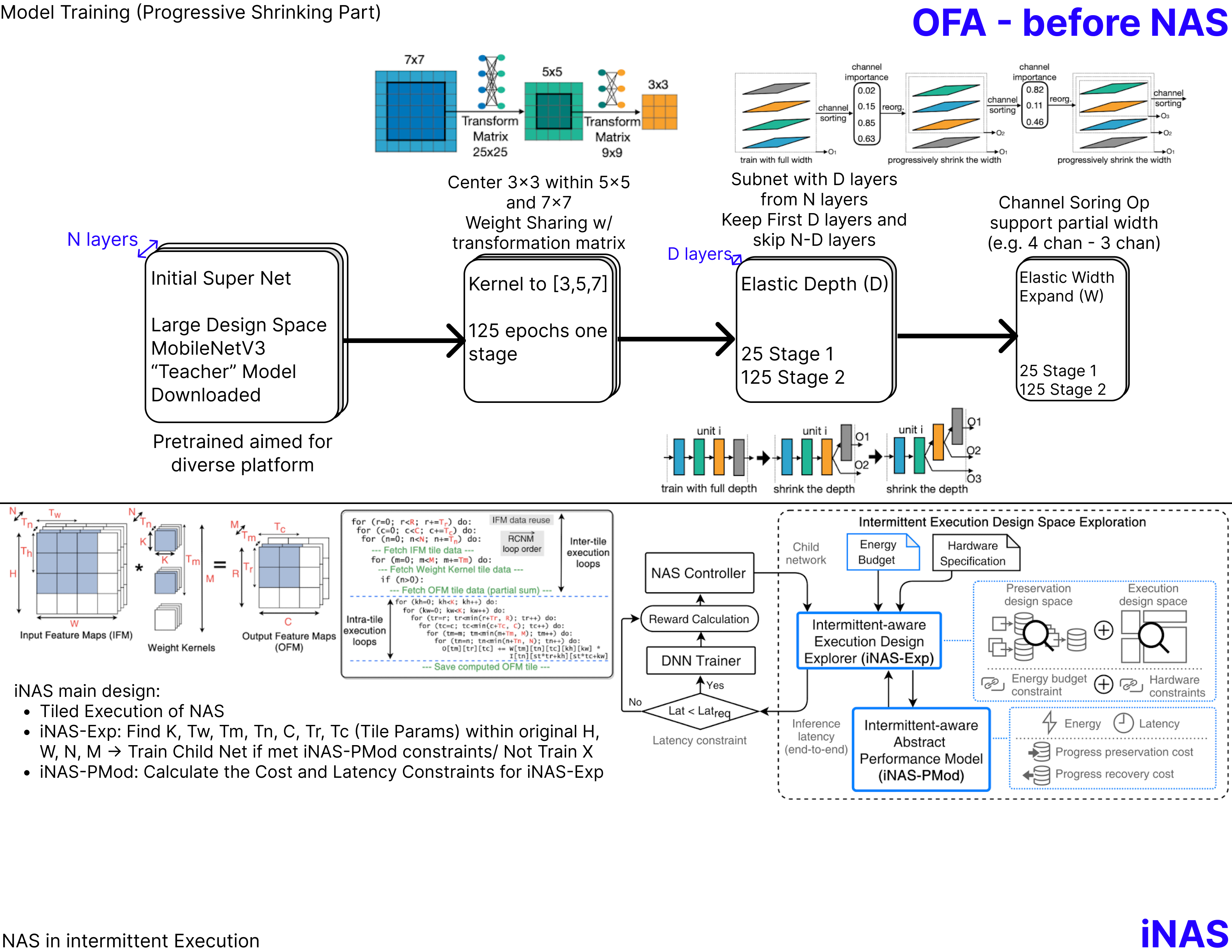}
  \caption{State-of-art baseline Appraoch - OFA and iNAS}
  \label{baseline}
\end{figure}

Executing DNN inferencing tasks often requires consecutive power to accumulate results over time. To execute Intermittently means that the model has to be broken down into multiple power cycles. Through data preservation and data recovery between a volatile memory (VM) and Non-Volatile Memory (NVM), the model can be executed safely within the deadline requirement. The common methodology uses Tiled Deep Neural Net computation to perform inter-level and intra-level Deep Neural Net Computation. The inner level consists of a single vector math operation typically supported by modern ultra-low power off-the-shelf MCU \cite{web1}. Therefore, to compute this single vector math operation into the most inner-level loop (intra-level loop), we can directly use a vector multiply-accumulate command \verb|LEACMD__MAC|. A Direct Memory Access handler is then configured with Serial Peripheral Interface (SPI) to transfer a single byte of input/output data.

DNNs require significant computational resources due to their loop-heavy and data-intensive computations. Therefore, their deployment on edge devices with limited resources requires execution design considerations such as data reuse schemes. Data reuse schemes primarily involve loop tile size and loop order. Loop tile size determines the size of the data blocks used in computations, and loop order refers to the order in which these blocks are processed. These schemes aim to minimize the amount of data movement between non-volatile memory (NVM) and volatile memory (VM) during successive computations, thus reducing the inference latency.

Overall, careful consideration of DNN architecture and execution design is crucial in optimizing DNN inference accuracy and latency on resource-constrained edge devices.

\subsection{Neural Architecture Search (NAS)}
 In the field of deep learning, hyperparameters refer to parameters that cannot be directly learned from training data. These parameters must be set before training, and can greatly affect the performance of a deep neural network (DNN). Examples of hyperparameters include the depth, width, height, kernel size, and tile size of the network \cite{Real2018RegularizedEF}.

One approach to optimizing hyperparameters for different platforms or hardware is to conduct a search over the parameter space. This is known as Neural Architecture Search (NAS). In NAS, the goal is to find the best-performing architecture for a given task, while minimizing the number of parameters and computational cost Fig.\ref{baseline}.

Although automatic neural network architecture prediction research dates back to the 1980s, interest has recently grown due to the success of deep neural networks in AI fields. As architectures become deeper, the search space grows exponentially, making the search process difficult. There are currently two main directions in searching for an architecture: using reinforcement learning or applying evolutionary algorithms. The NAS framework is an iterative process that generates a new child network, trains it on a held-out data set, and uses the resulting accuracy as a reward signal for the next iteration. The process stops when the controller has converged for maximum accuracy or when the child network achieves the required accuracy. While the automatically searched network architectures can achieve accuracy close to human-invented architectures, there are two main challenges that must be addressed. First, the searching process is inefficient and time-consuming. Second, the generated neural architectures can be slow and complex, making them unsuitable for real-time AI applications that require fast inference speed.

Traditional NAS tends to focus on maximizing data reuse among different layers, without taking into account resource constraints. However, with the rise of TinyML (machine learning on ultra-low power devices), there has been an increasing focus on incorporating resource constraints into NAS. TinyNAS and TinyEngine \cite{Lin2020MCUNetTD} \cite{Lin2021MCUNetV2MP} are examples of NAS models that are designed specifically for resource-constrained environments.

In TinyML \cite{Lin2020MCUNetTD}, memory constraints are a critical factor in optimizing network performance. Modern NAS models can adapt to different layer sizes through memory scheduling, maximizing memory usage under constraints and increasing data reuse to reduce runtime overhead. \cite{Lin2020MCUNetTD} By efficiently scheduling memory usage, these models can achieve high accuracy while operating within the constraints of a given platform or hardware.

\subsection{Intermittent-aware NAS}
Intermittent systems, as described in the passage, operate by storing energy in a buffer, which is typically a capacitor. When the buffered energy reaches a preset threshold, the system is powered on, and when the energy buffer is depleted or reaches a low threshold, the system is powered off. The amount of energy available to the system, or the energy budget, is defined by the buffered energy, and this determines how long the system can stay on during each power cycle. The duration of the power on and off cycles is dependent on the energy budget and consumption, as well as the recharge time of the buffer, which is determined by the size of the buffer and the strength of the power supply Fig.\ref{baseline}.

In order to overcome power failures and prevent data loss, intermittent execution preserves progress by saving progress indicators and outputs to non-volatile memory (NVM). When power is restored, progress recovery is performed by restarting the system and retrieving the saved progress indicators and input data. Interrupted atomic operations are re-executed upon power resumption to ensure the system continues from where it left off.

Task-based execution models are commonly used to partition applications into atomic tasks, where the energy budget must satisfy the energy consumption of each task, and progress preservation is performed after task completion. In the case of intermittent DNN inference, progress indicators are represented by inter-tile loop indices, which are preserved along with tile output data as part of progress preservation during inference. Batch preservation is used to reduce the progress preservation overhead, by retaining a batch of computation outputs in VM and preserving the S inference outputs together. The batch size S needs to be appropriately set to balance preservation overhead and progress loss due to power failure. Additional VM may be required to retain S computation outputs.

Fitting the generated child networks from NAS to intermittent execution often relates to how to design a safe preservation and recovery mechanism and satisfies the latency requirement and hardware requirement. Recent work iNAS \cite{iNAS} provided a complete design to incorporate intermittent execution in the phase of NAS. It proposed a NAS controller that implements Intermittent-aware Execution Design Explorer (iNAS-Exp) and Intermittent-aware Abstract Performance Model (iNAS-PMod) to derive a feasible solution space and test against latency requirements and hardware constraints. iNAS-Exp uses progress preservation cost and progress recovery cost calculated by iNAS-PMod to determine preservation design space and execution design space for the NAS. This system satisfies the deadline by mapping tiles DNN tasks: computation, progress preservation, progress recovery, and charging according to power cycles, and only performs NAS once constraints are met. Therefore, its DNN can be safely executed on the intermittent system.

\section{Methodology}

We identify the question as improving the current intermittent DNN execution to achieve better accuracy and less latency for the metrics of the Real-time System. The Approaches can be broken down further into the following two aspects: (1) Conducting Real-Time Tasks Scheduling \cite{9378771} with the DNN inferencing tasks on intermittent systems. (2) Optimizing the baseline NAS \cite{iNAS} algorithm to be more light-weighted and memory efficient and is able to generate a model with higher accuracy within the same degree of design freedom. 

\subsection{Realtime Schedulability Analysis}
In the context of real-time systems, it is essential to ensure that tasks meet their deadlines and are executed within a specified time frame. The worst-case execution time (WCET) is a measure of the maximum time required for a task to complete, and it is used to determine whether a real-time system can meet its deadlines.

The original intermittent deep neural network (DNN) inference did not include any form of schedulability analysis for DNN tasks. This means that we cannot determine the reliability of the system, including the worst-case execution time, which is a crucial factor for real-time systems.

Our approach fills this gap by providing a scheduling framework specifically designed for intermittent deep neural net tasks. By incorporating schedulability analysis into the system, we can schedule a larger workset and reduce scheduling overhead. This, in turn, will accelerate deep inference tasks, making them more efficient and reliable

\begin{figure}[htbp]
  \centering
  \includegraphics[width=0.95\linewidth]{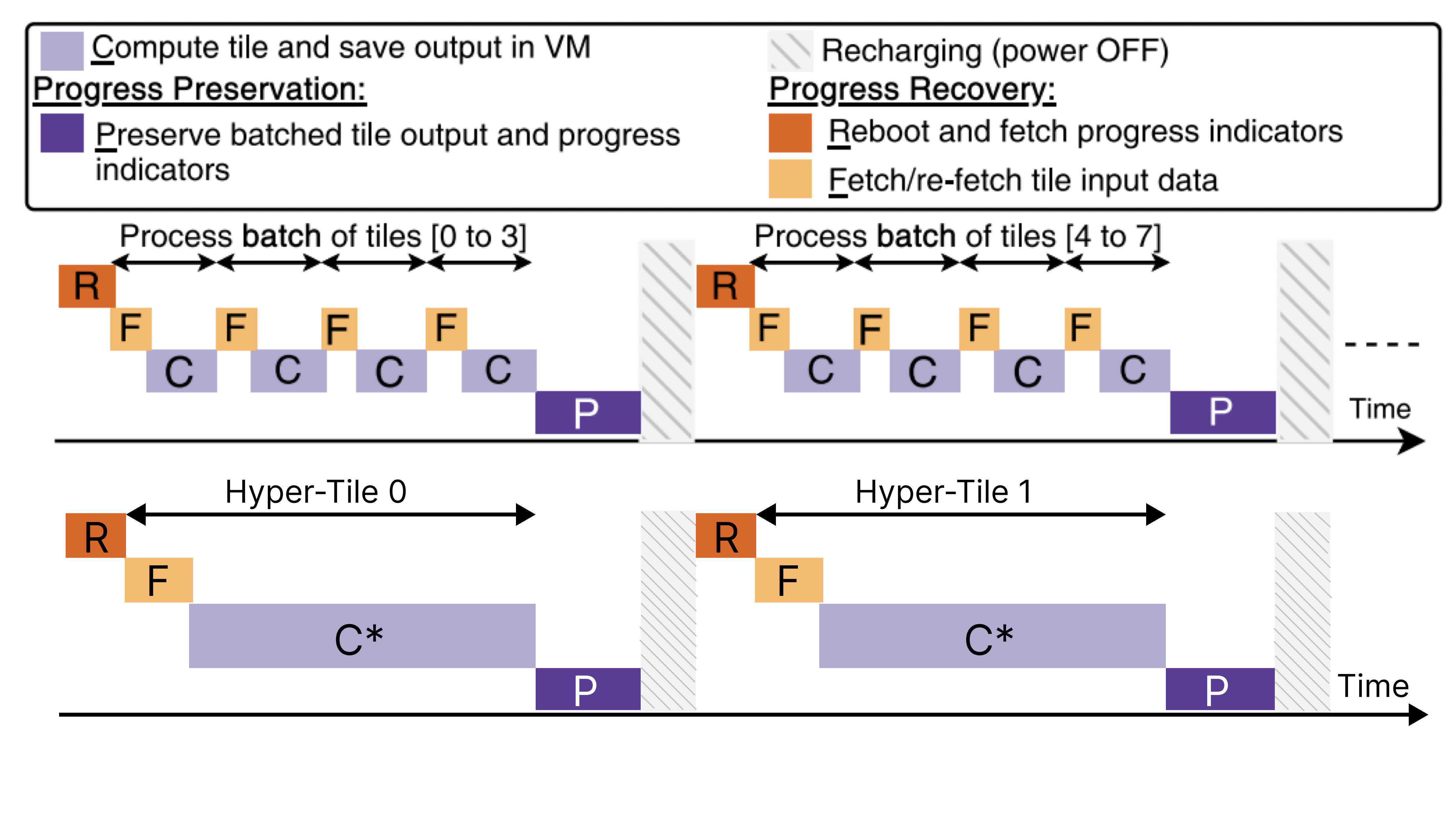}
  \caption{Scheduling Optimization}
  \label{sched}
\end{figure}

Our scheduling framework will allow for better management of DNN tasks, ensuring that they are executed within their deadlines and without any unexpected delays. The framework will also enable us to optimize the allocation of system resources, such as memory and processing power, to improve the overall performance of the system.

By providing a scheduling framework for intermittent deep neural net tasks, our project will contribute to the development of more reliable and efficient real-time systems, with potential applications in areas such as autonomous vehicles, medical devices, and industrial automation.

\begin{figure}[htbp]
  \centering
  \includegraphics[width=0.95\linewidth]{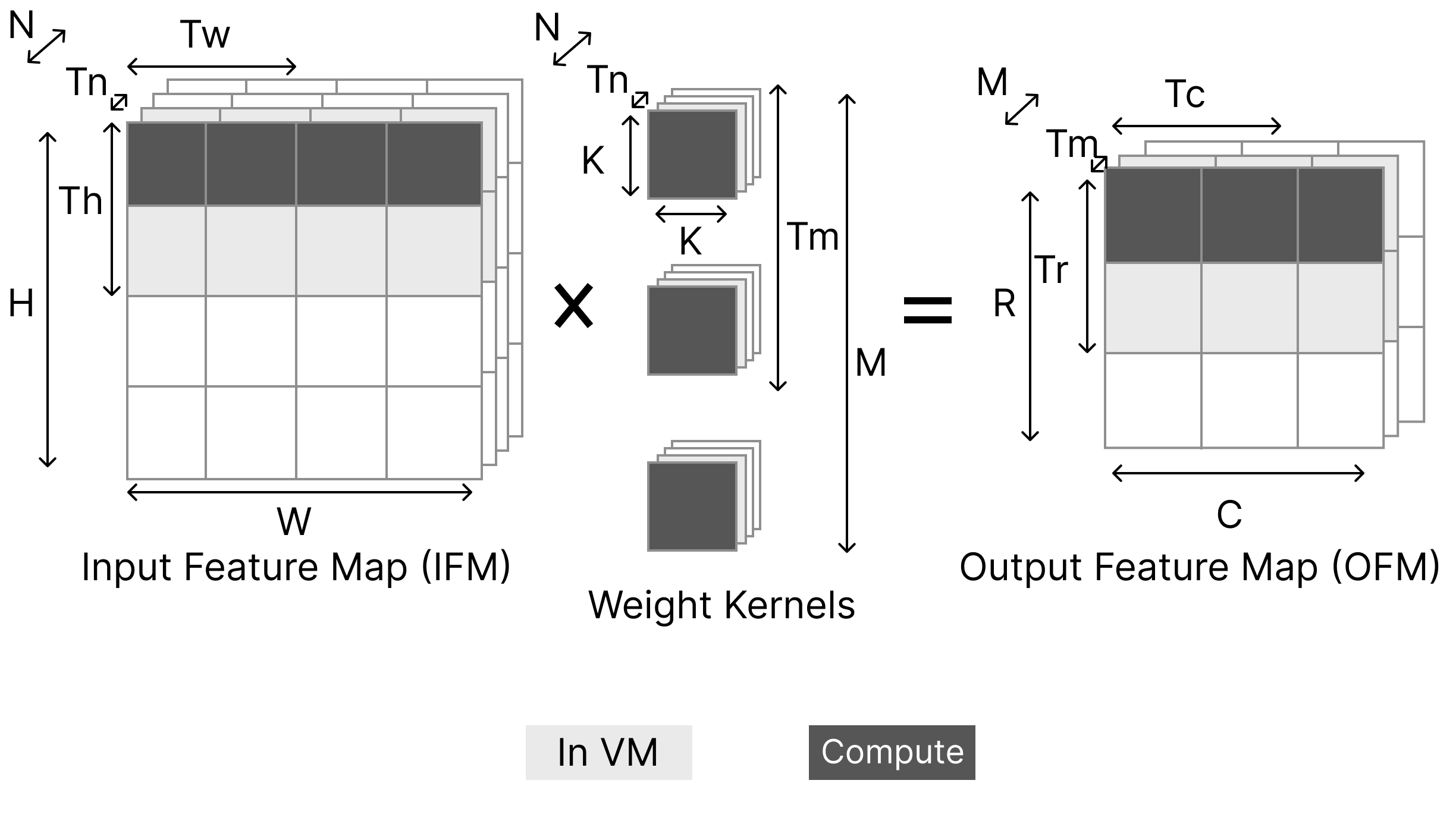}
  \caption{More efficient Tiled DNN: Increase granularity due to a larger budget}
  \label{r-dnn}
\end{figure}

The Vanilla iNAS approach Fig.\ref{dnn} is designed to generate neural network models that are highly granulized for extremely small memory size, allowing for less data to be stored in Volatile Memory. However, this approach also requires frequent data fetching and results in higher fetch overhead and less computation, which can impact overall performance.

In contrast, the Accelerated insNAS approach Fig.\ref{r-dnn} aims to increase efficiency and throughput by adopting a large batch size, known as hyper-tile. This approach allows for more data to be processed in one power cycle, which results in more computation and, ultimately, more efficiency. In practice, this approach has been shown to enable the creation of magnitude larger CNNs with only a small increase in the amount of SRAM required.

Overall, the accelerated insNAS approach improves upon the Vanilla iNAS approach by balancing the need for highly granulized models with the requirement for more efficient computation and less fetch overhead. By increasing the batch size and processing more data at once, this approach is able to achieve higher throughput while minimizing the impact on memory requirements.

\subsection{Accelerating NAS with TinyML}
\textbf{Compile the float16 data to the runtime binary (header file) to optimize for data-fetch overhead}. Compiling float16 data to the runtime binary (header file) is a technique used to optimize neural network performance. Float16 data uses 16 bits to represent floating-point numbers, which can reduce the memory requirements of neural networks and speed up computation.

In order to optimize for data-fetch overhead, float16 data can be compiled to the runtime binary (header file) of the neural network. This allows the data to be loaded more efficiently, as it is stored in a format that can be directly accessed by the neural network. This approach can improve the performance of neural networks, especially in scenarios where the data needs to be loaded quickly and efficiently.

\textbf{Knowledge distillation} is a process where a larger, more complex model (known as the teacher model) is used to train a smaller, simpler model (known as the student model). This process can be used to improve the accuracy and efficiency of models, especially in scenarios where computational resources are limited.

\begin{figure}[htbp]
  \centering
  \includegraphics[width=0.95\linewidth]{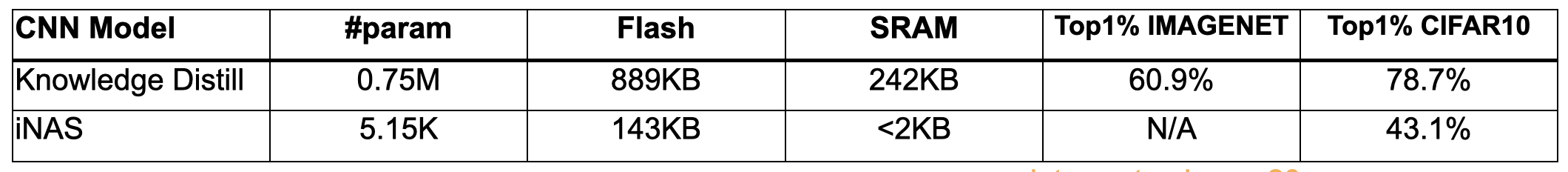}
  \caption{Weight Sharing Model after Knowledge Distillation}
  \label{table1}
\end{figure}

One approach to knowledge distillation is \textbf{progressive shrinking}, where the teacher model is reduced in three perspectives: Width, Kernel and Channel. The student model is trained on each intermediate version of the teacher model. This approach has been shown to be effective in reducing the gap between the accuracy of the teacher and student models Fig.\ref{table1}.

Another approach to knowledge distillation is \textbf{TinyNAS+TinyEngine Co-design}. In this approach, a small and efficient neural architecture search (NAS) algorithm is used to search for an optimal network architecture, and this architecture is then used to train a student model using knowledge distillation. This co-design approach can lead to more efficient and accurate models compared to traditional knowledge distillation methods.

\textbf{Weights Sharing: Store the Final Weights After NAS to csv files and try with layer-by-layer execution}: Following Fig.\ref{table1}, Weights sharing is a technique used in neural network compression to reduce the number of parameters in a model. This technique involves sharing the weights between different layers in a neural network, thereby reducing the overall number of parameters in the model. One approach to weights sharing is to store the final weights after NAS to csv files and try with layer-by-layer execution. This involves training a neural network using a NAS algorithm, and then saving the final weights to a csv file. The csv file can then be used to initialize the weights of a new model, which can be trained using a layer-by-layer execution approach. This approach can be more efficient than traditional training approaches, as it reduces the number of parameters that need to be trained.

\section{Implementation}

\subsection{iNAS Baseline Study}

\begin{figure}[htbp]
  \centering
  \includegraphics[width=0.95\linewidth]{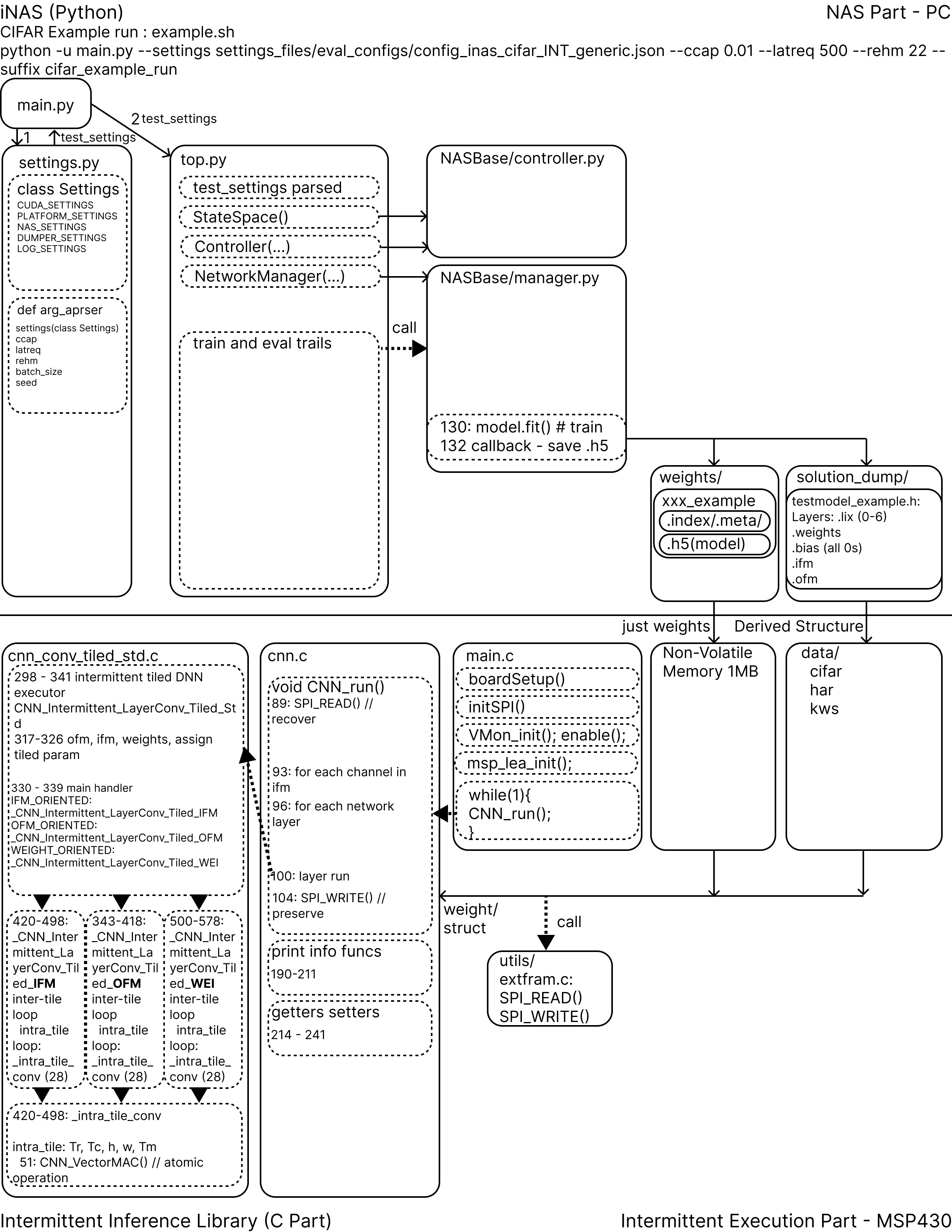}
  \caption{iNAS Workflow Design}
  \label{iNAS}
\end{figure}

The iNAS system has two important components: iNAS-Exp and iNAS-PMod. These components work with traditional components such as the NAS controller and the DNN trainer. The NAS controller uses a recurrent neural network to explore the architecture space and reinforcement learning to update the controller. Each iteration of the NAS controller generates a child network that is passed to the iNAS-Exp to find the intermittent execution design that minimizes the end-to-end inference latency. The iNAS-Exp performs a preservation and execution design space exploration on the child network, while ensuring the feasibility of a candidate design.

For each candidate design in the search space, iNAS-PMod is used to estimate the energy consumption per power cycle and the end-to-end inference latency. These estimates are used to determine if a candidate design can execute within the energy budget and without violating the latency requirement. If the lowest achievable end-to-end inference latency by any of the feasible candidate designs is lower than the latency requirement, the DNN trainer is used to train and obtain the accuracy of the child network.

Then, the accuracy and the lowest achievable end-to-end inference latency are used as a reward signal to update the NAS controller. The reward signal is formulated similarly to an existing HW-NAS, where the reward is equal to the accuracy minus the exponential moving average of previous feasible child networks' accuracies, plus the latency divided by the latency requirement. This process is repeated for a specified number of iterations, guiding the search towards a network model that is accurate, safe, and efficient.

The resulting solution of iNAS is the generated child network with the highest accuracy and its corresponding intermittent execution design which has the lowest end-to-end inference latency. The total search time increases linearly with the number of generated feasible child networks. Although the architecture search space used in iNAS is relatively simple, more complex architecture search spaces can be easily integrated into iNAS.

\subsection{Accelerated intermittent NAS}

We have conducted everal optimization ways that can be implemented on the iNAS framework described above to improve its performance and efficiency:

\begin{figure}[htbp]
  \centering
  \includegraphics[width=0.95\linewidth]{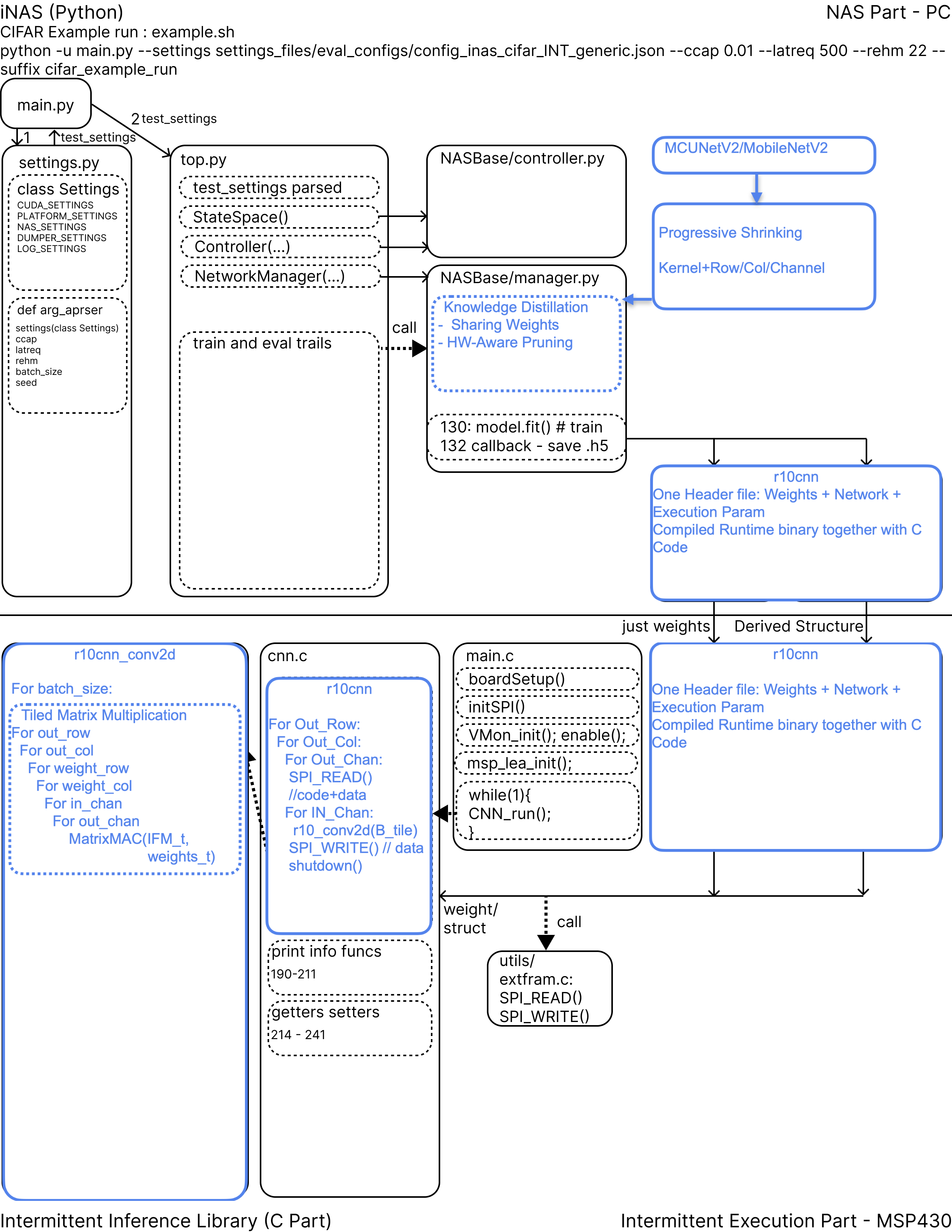}
  \caption{Accelerated intermittent NAS Workflow Design}
  \label{aiNAS}
\end{figure}

\begin{itemize}
\item Knowledge Distillation from the teacher super net. We introduced our teacher model to be as large as possible before the progressive shrinking (e.g. MobileNet, MCUNetV2). Before running the NAS itself, the program performs a knowledge distillation sequence to prune more than \verb|80%| of the parameters and then share the weights with the least accuracy decrease based on the biases during the shrinking. The derived network is pruned and shrunk with three dimensions: Kernel (Weights), Row/Col (Depth) and Channels. We then feed the network to the iNAS NAS Controller to perform the search based on the configurations given different hardware.

\item Use more sophisticated NAS controllers: iNAS currently uses a NAS controller based on a recurrent neural network. More sophisticated controllers, such as those based on evolutionary algorithms or Bayesian optimization, can be integrated into the framework to further improve the search process.

\item After the NAS, The program uses a solution dumper to generate the C Code that is used to run on Micro-controller. We generate a single header file that contains all the weights information and CNN structure so that we can fetch them all at once runtime. To be able to compile the C Code together, some external hardware libraries are needed based on the specific platform you are using. For example, in our case study we have used the MSP430 library (DSPLib)\cite{web2} given by Texas Instruments and produced the runtime code with specific functionality like SPIRead and SPIWrite. 
\end{itemize}

\subsection{Discussion of using hyer-tile}

The benefit of adopting the large batch size, or hyper-tile, in the Accelerated insNAS approach is that it leads to increased efficiency and throughput. By processing more data in a single power cycle, more computation can be done, which ultimately leads to higher efficiency. This is in contrast to the Vanilla iNAS approach, which aimed to achieve highly granulized models with smaller memory requirements.

In practice, the use of hyper-tile has been shown to enable the creation of magnitude larger CNNs with only a small increase in the amount of SRAM required. This means that the Accelerated insNAS approach is able to achieve higher levels of efficiency without significantly increasing the memory requirements of the system. This is particularly beneficial for applications that require high throughput and efficient computation, such as real-time image processing and video analysis.

Overall, the use of hyper-tile in the Accelerated insNAS approach represents an important optimization strategy that allows for the creation of larger and more complex CNNs while still maintaining high levels of efficiency and minimizing the impact on memory requirements. By achieving a better balance between model complexity and computational efficiency, this approach is able to improve upon the limitations of the Vanilla iNAS approach and provide a more practical and effective solution for real-world applications.

\section{Evaluation}

\subsection{Hardware Setup - MSP430}
The MSP430FR5994 is a microcontroller developed by Texas Instruments that is commonly used in low-power embedded systems. It features 2KB of SRAM, which is a volatile memory that can be quickly accessed but loses its contents when power is removed. Additionally, it has 256KB of FRAM, which is a non-volatile memory that can retain its contents even when power is removed, making it suitable for storing program code or critical data.

In addition to the on-board memory, the MSP430FR5994 can also interface with external memory devices, such as the Excelon LP 8-Mbit SPI F-RAM. This allows for additional storage capacity to be added to the system, while still maintaining the low power consumption and fast access times of the FRAM technology.

The energy source for the MSP430FR5994 is typically provided by a PC or other external power supply. To enable intermittent execution, a voltage monitor can be used to detect when the power supply voltage drops below a certain threshold, indicating a power loss event. This trigger can then be used to save the current state of the system to non-volatile memory, such as the FRAM, before shutting down the system to conserve energy.

The reason to use the MSP430FR5994 is that it is a versatile and low-power microcontroller that is well-suited for applications where power consumption and non-volatile memory storage are important considerations. Its ability to interface with external memory devices and support intermittent execution makes it a popular choice for a wide range of embedded systems.

\begin{figure}[htbp]
  \centering
  \includegraphics[width=0.95\linewidth]{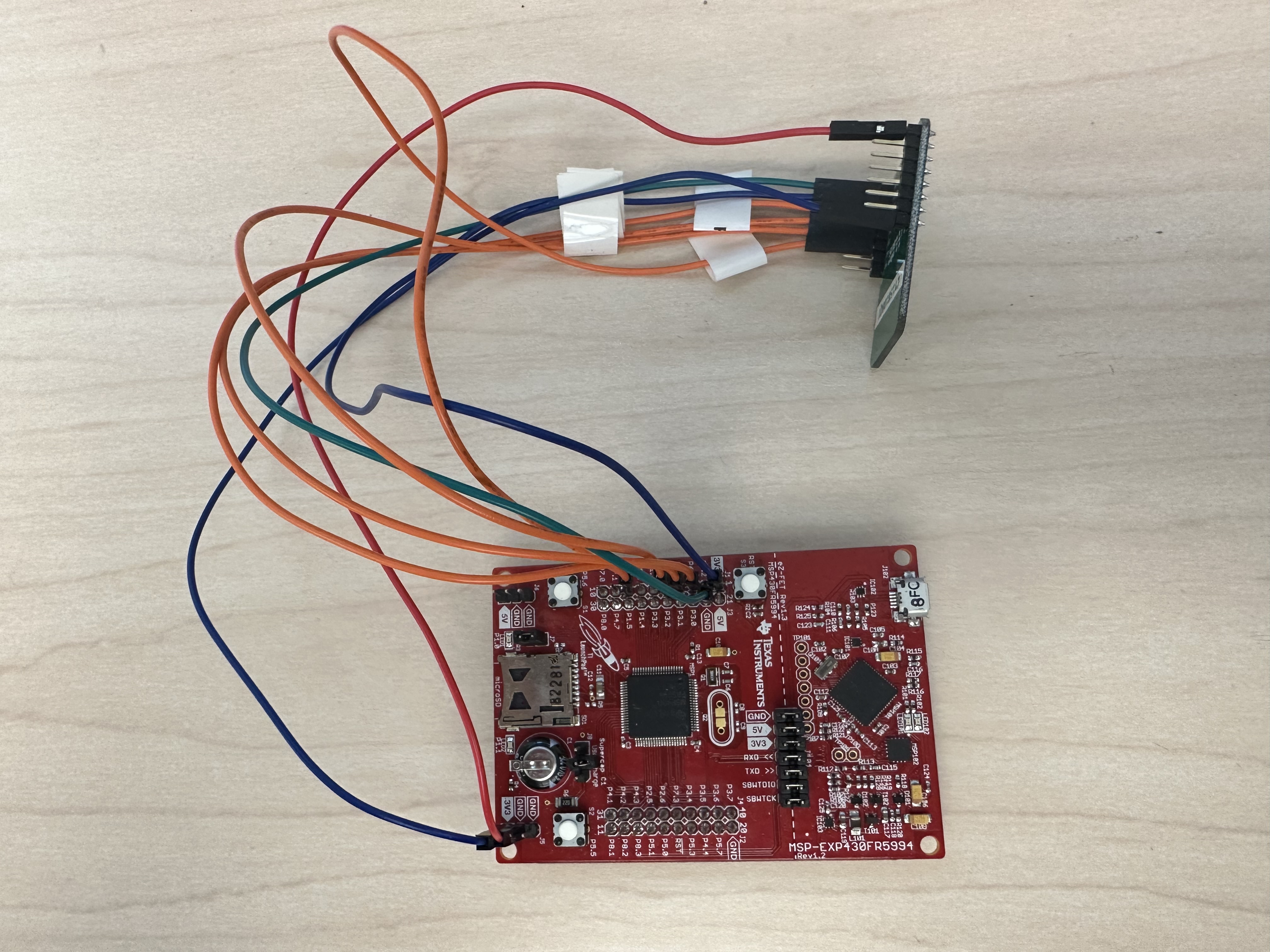}
  \caption{Experiment Hardware Setup}
  \label{msp430}
\end{figure}

\subsection{Dataset Setup}
\begin{figure}[htbp]
  \centering
  \includegraphics[width=0.95\linewidth]{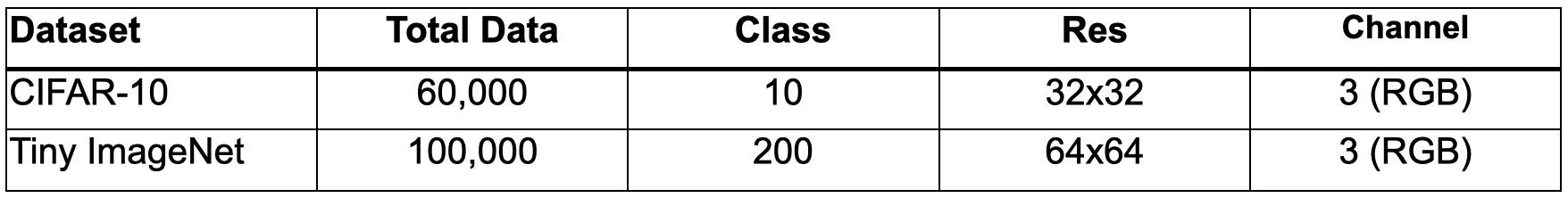}
  \caption{Experiment Dataset Chart}
  \label{data}
\end{figure}

We have tested all baseline models and our implementation on two dataset Fig.\ref{data}: CIFAR-10 and Tiny Imagenet. The reason to choose these two datasets is that we want to test out how good each candidate can perofrm under a huge scalibility different between the small light-weighted dataset and some actual life scenario. 

CIFAR-10 consists of 60,000 32x32 color images in 10 classes, with 6,000 images per class. The classes include airplanes, automobiles, birds, cats, deer, dogs, frogs, horses, ships, and trucks. The dataset is divided into 50,000 training images and 10,000 testing images. 

The Tiny ImageNet dataset is a smaller version of the original ImageNet dataset, which is a widely used benchmark dataset in machine learning for image classification tasks. The Tiny ImageNet dataset consists of 200 object classes, with each class having 500 training images, 50 validation images, and 50 test images. The dataset contains 100,000 training images, 10,000 validation images, and 10,000 test images in total. Each image in the Tiny ImageNet dataset is 64x64 pixels in size, which is much smaller than the original ImageNet dataset. The object classes in the Tiny ImageNet dataset are a subset of the object classes in the original ImageNet dataset, and they were chosen to be representative of a wide range of object categories, including animals, vehicles, and household items.

\subsection{NAS Setup}

\begin{figure}[htbp]
  \centering
  \includegraphics[width=0.95\linewidth]{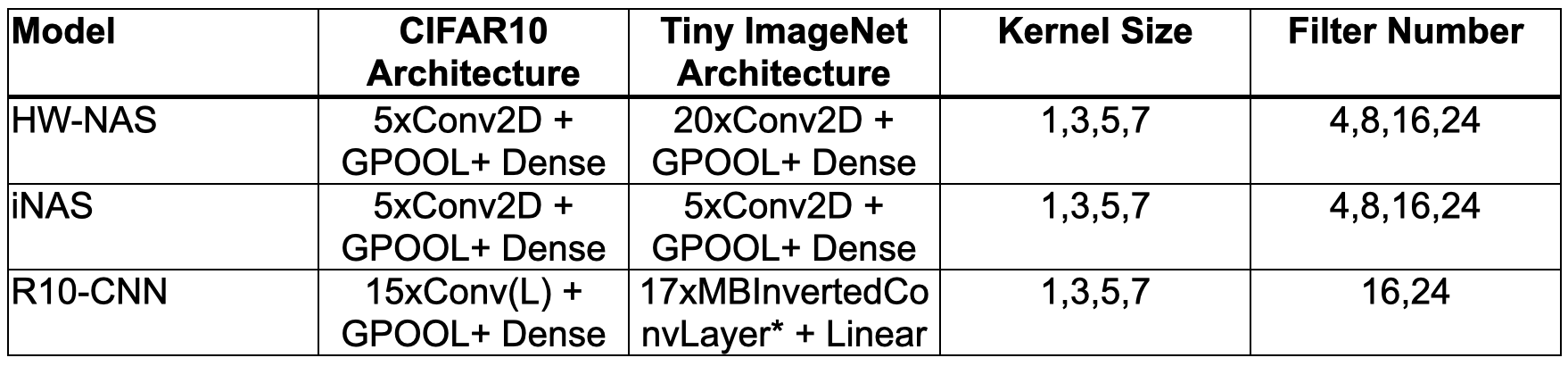}
  \caption{Experiment NAS Setup Chart. *MBInvertedConvLayer: 3xConv2D + 3xBatchNormalize2D + ConvLayer + 2xReLU}
  \label{nas}
\end{figure}

Our final setup is regarding the NAS Controller setup Fig.\ref{nas} to fix a structure for each of the models. The baselines model structure is directly obtained from their relevant paper and our NAS setup is derived based on the model we have chossesn as the teacher mode. In this experiment, we have chosen the MCUNetV1-Int2 \cite{Lin2020MCUNetTD} which uses the half float precision (float16) and has the base Top \verb|1%| accuracy for Imagenet at \verb|60.9%|. The problem for this net is that it cannot be executed on a intermittent power budget and the runtime memory is around 248KB. However the MSP430 we have used only have a runtime SRAM at 2KB so that a NAS Controller was introduced to further schedule it in the intermittent power cycle with a peak runtime memory at 2KB

\subsection{Accuracy Gain}

\begin{figure}[htbp]
  \centering
  \includegraphics[width=0.95\linewidth]{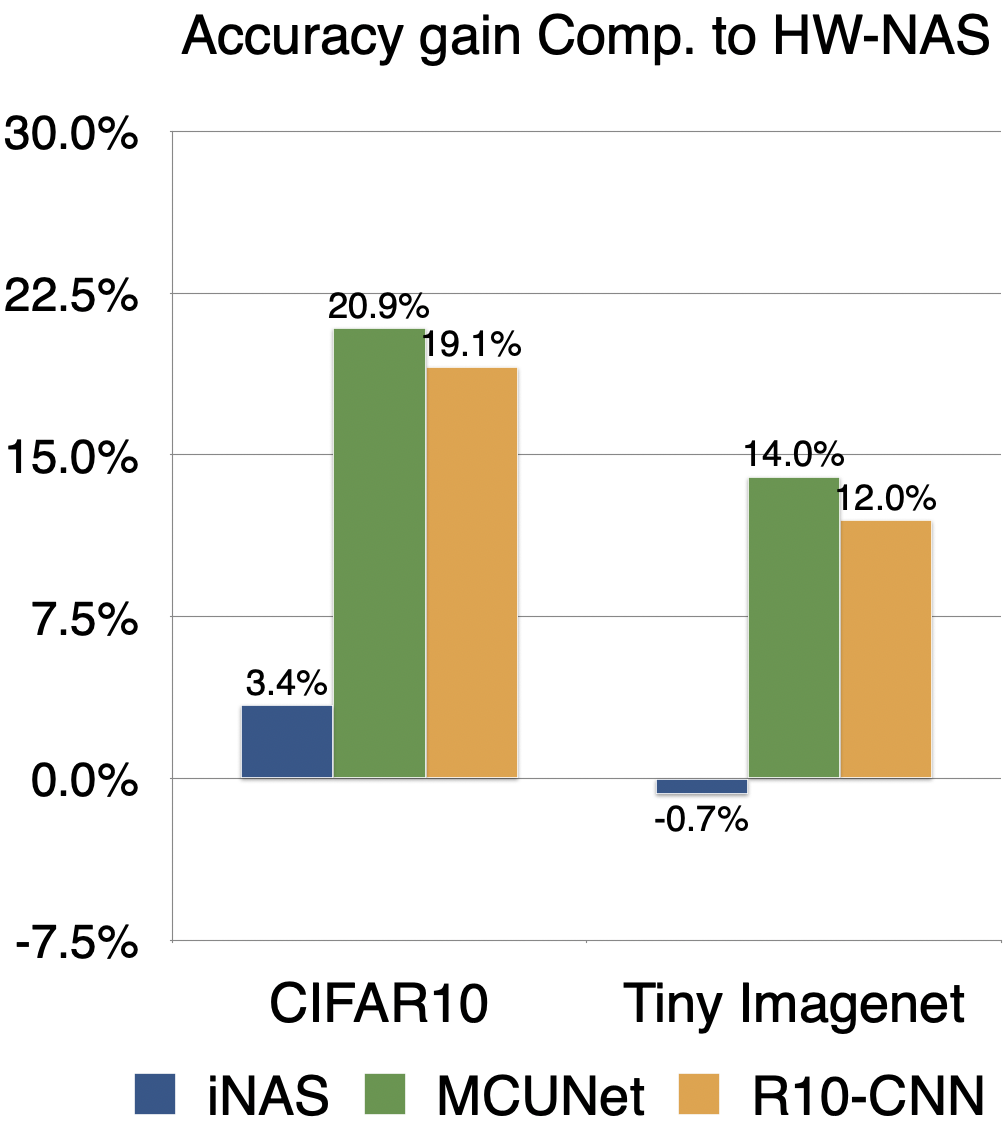}
  \caption{Accuracy Gain}
  \label{eval-acc}
\end{figure}

One of our main contributions is to increase the accuracy of the current baseline models that is runnable on MSP430 greatly and make dataset like Imagenetto be inferrable under the intermittent power execution. In our Fig.\ref{eval-acc} we have reported the accuracy increase comapred to the original Hardware-NAS and reached a \verb|19.1%| increase in CIFAR-10 dataset and a \verb|12.0%| in Tiny Imagenet Dataset. The result can be further analyzed by comapring to the iNAS implementation, which reached a comparable accuracy comapred to HW-NAS. All of the accuracy metrics are conducted on PC platform with Only CPU inference enabled.

\subsection{Peak Runtime Memory}

\begin{figure}[htbp]
  \centering
  \includegraphics[width=0.95\linewidth]{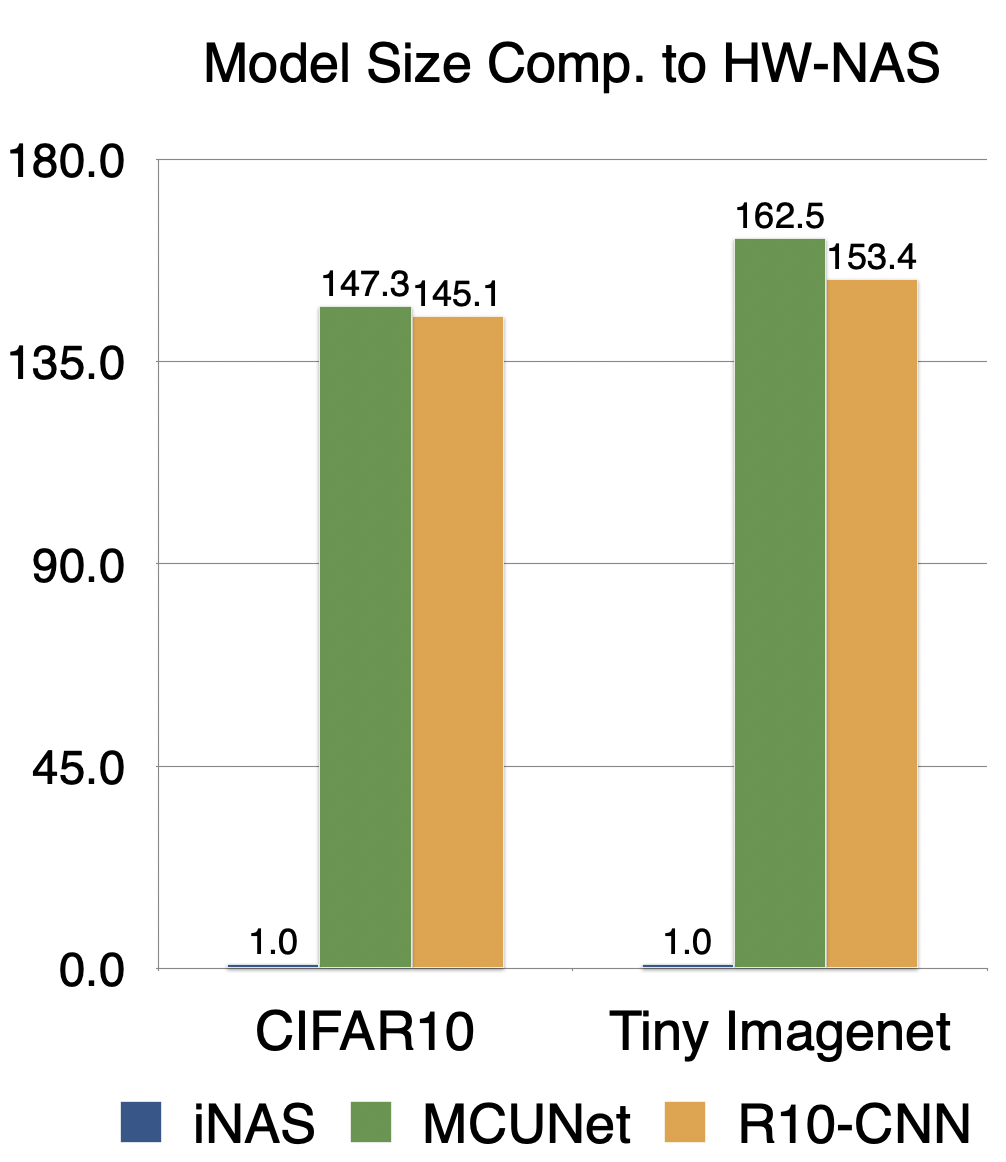}
  \caption{Model Size comparison}
  \label{eval-modelsize}
\end{figure}

Fig.\ref{eval-lat} reported the peak runtime memory of the models. In the figure we can see our model has reached a 1.33x times of runtime memory (\verb|33%| increase) on the CIFAR-10. For Tiny-Imagenet our solution only reaches 1.27x memory increase (\verb|27%|) compared to baseline HW-NAS. However, comapred to runtime memory of MCUNet which is at 248KB for CIFAR-10, our model is able to run a much larger (145X times)Fig.\ref{eval-modelsize} network using a comaparable peak runtime memory capped at 2KB. 

\begin{figure}[htbp]
  \centering
  \includegraphics[width=0.95\linewidth]{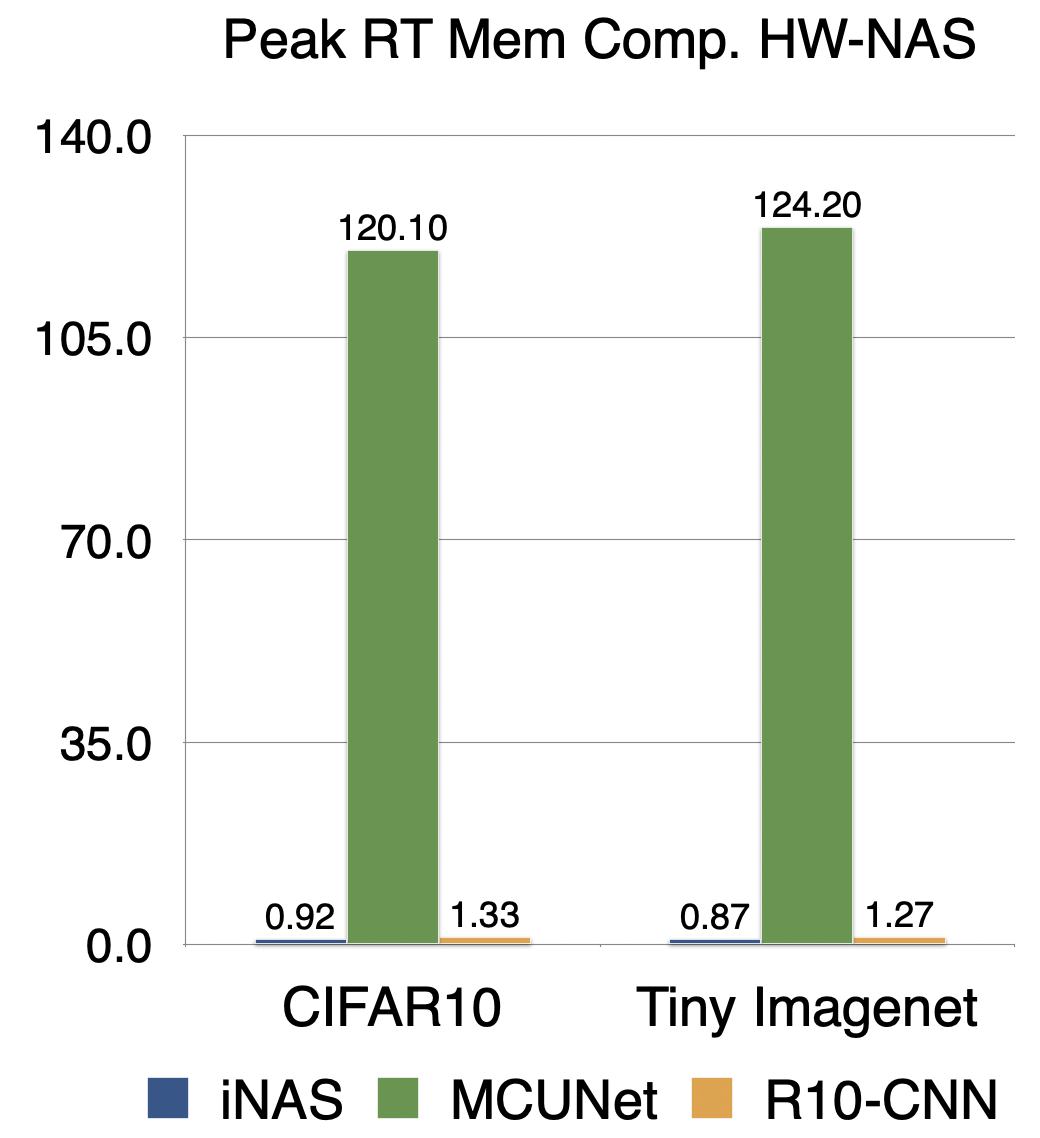}
  \caption{Peak Runtime Memory}
  \label{eval-peakmem}
\end{figure}

\subsection{Latency}

\begin{figure}[htbp]
  \centering
  \includegraphics[width=0.95\linewidth]{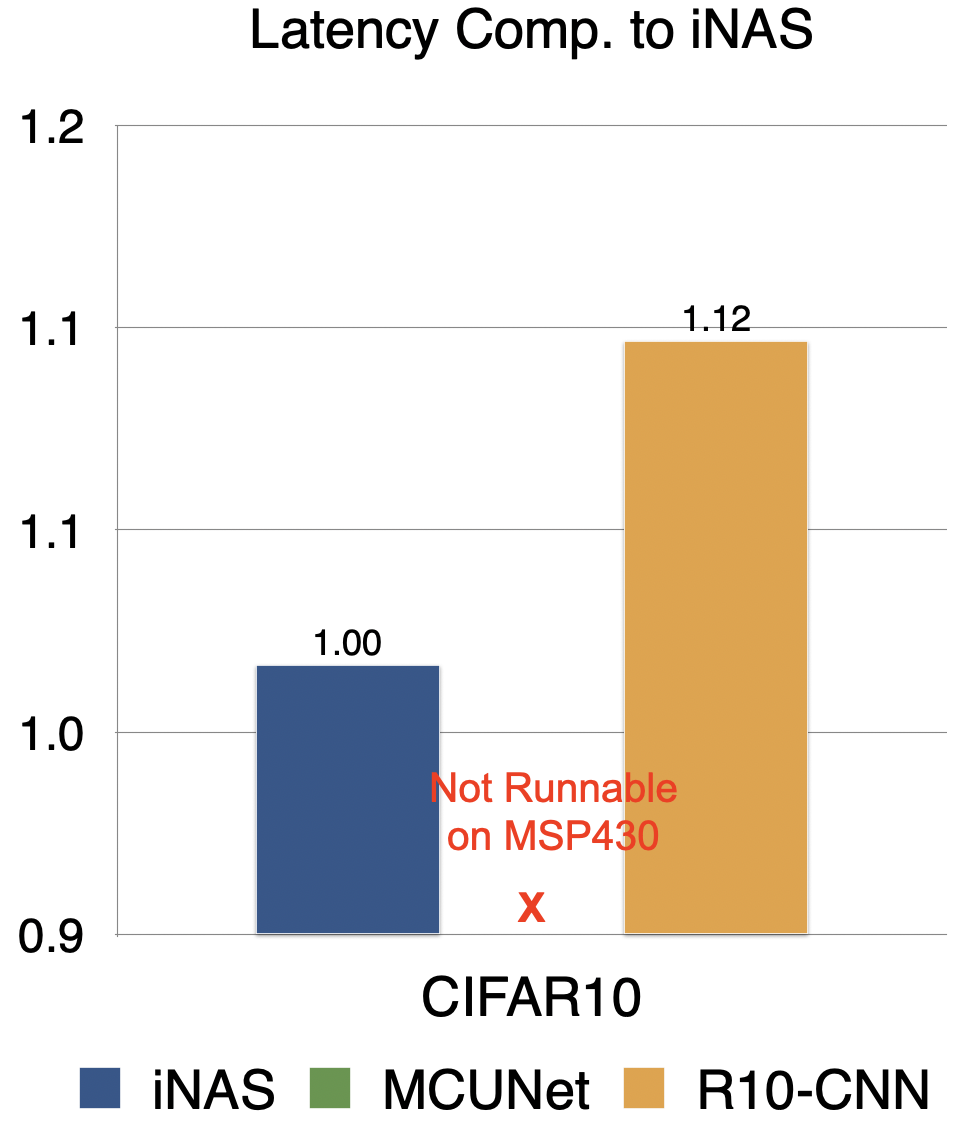}
  \caption{Inference Latency}
  \label{eval-lat}
\end{figure}

Our final experiments are to deploy the models onto the MSP430 and test for the runtime latency using the CIFAR-10. Compared to iNAS, MCUNet is not able to run on the MSP430 itself due to the extremely large runtime memory and our model is runnable but at a \verb|12%| increase of latency comapred to iNAS Fig.\ref{eval-lat}.

\section{Discussion}

We claimed the following based on our models and data from the evaluation: Make 145X larger size model runnable 
for \verb|41%| increase of SRAM while maintaining comparable latency at \verb|12%| overhead. The tradeoff is made between accuracy and latency. We are able to make a 0.75 milion parameters netowrk to be runnable under 2KB and followed the intermittent execution pattern. We claimed that we are the first to build a life scenario usable (Trained and Inferenced under Tiny Imagenet) model that can reach more than \verb|60%| of accuracy. We admitted iNAS is more optimal regarding the smaller network under the intermittent execution, but such network can be further improved by our model and achieve a higher accuracy by adopting the ideas of knwoeldge distillation and progressive shrinking.

\section{Future Research}

As the new hardware for ultra-low power devices are made, we can then conducted the experiment using a much larger RAM intermittent device like Ambiq Apollo 4 Blue Plus~\cite{a4p}. It has a volatile SRAM at 2.75MB and Non-Volatile Memory MRAM at 2MB. It also supports an ARM 32Bit SoC that can perform the full precision inference on board. Therefore, we can analyze the following questions: (1)Given sufficiently large (one layer) SRAM, is tiling still needed? (2) If we can run  multiple DNN together, can we further study for the tradeoff of latency and accuracy~\cite{zhang2024mii}?

\bibliographystyle{plain}
\bibliography{ref}

\end{document}